\documentclass[preprint,aps]{revtex4}
\usepackage{graphicx,epsfig,subfig,dcolumn,bm,mathrsfs,amsmath}
\begin{document}

\title{A Dissipative-Particle-Dynamics Model for Simulating Dynamics
  of Charged Colloid}

\author{Jiajia Zhou}
\email[]{zhou@uni-mainz.de}
\altaffiliation{}
\affiliation{Institut f\"ur Physik, Johannes Gutenberg-Universit\"at Mainz\\
 Staudingerweg 7, D55099 Mainz, Germany} 
\author{Friederike Schmid}
\email[]{friederike.schmid@uni-mainz.de}
\affiliation{Institut f\"ur Physik, Johannes Gutenberg-Universit\"at Mainz\\
 Staudingerweg 7, D55099 Mainz, Germany} 

\begin{abstract}
A mesoscopic colloid model is developed in which a spherical colloid is represented by many interacting sites on its surface. 
The hydrodynamic interactions with thermal fluctuations are taken accounts in full using Dissipative Particle Dynamics, and the electrostatic interactions are simulated using Particle-Particle-Particle Mesh method. 
This new model is applied to investigate the electrophoretic mobility of a charged colloid under an external electric field, and the influence of salt concentration and colloid charge are systematically studied.
The simulation results show good agreement with predictions from the electrokinetic theory.
\end{abstract}

\maketitle

\section{Introduction}
\label{sec:intro}

Colloidal dispersions have numerous applications in different fields such as chemistry, biology, medicine, and engineering \cite{RSS, Dhont}. 
In an aqueous solution, colloidal particles are often charged, either by ionization or dissociation of a surface group, or preferential adsorption of ions from the solution.
A good understanding of the dynamics of charged colloids is important from the fundamental physics point of view. 
Furthermore, such an understanding may also provide insights to improve the material properties of colloidal dispersions.
Theoretic studies of charged colloids are difficult because of the complexity of the system and various different interactions among the colloid, solvents, and microions. 

Molecular simulations can shed light on the dynamic phenomena of charged colloids in a well-defined model system.
Such studies are numerically challenging because two different types of long-range interactions are involved: the electrostatic and the hydrodynamic interactions.
In recent years, a number of coarse-grained simulation methods have been developed to address this class of problem. 
The general idea is to couple the explicit charges with a mesoscopic model for Navier-Stokes fluids. 
One of the examples is the coupling scheme developed by Ahlrichs and D\"unweg \cite{Ahlrichs1999}, which combines a Lattice-Boltzmann (LB) approach for the fluid and a continuum Molecular Dynamics model for the polymer chains.
This method has been applied to study the polyelectrolyte electrophoresis and successfully explained the maximum mobility in the oligomer range for flexible chains \cite{Grass2008, Grass2009, Grass2010}. 
Besides the Lattice-Boltzmann method \cite{Lobaskin2004, Lobaskin2004a, Lobaskin2007, Chatterji2005, Chatterji2007, Giupponi2011}, there are a few choices of the fluid model in the literature, such as the Direct Numerical Simulation (DNS) \cite{Tanaka2000, Nakayama2005, Kim2006, Nakayama2008}, Multi-Particle Collision Dynamics (MPCD) \cite{Malevanets1999, Gompper2009}, and Dissipative Particle Dynamics (DPD) \cite{Hoogerbrugge1992, Koelman1993, Espanol1995, Groot1997}.   
In this paper, we choose the DPD approach.  
DPD is a coarse-grained simulation method which is Galilean invariant and conserves momentum.
Since it is a particle-based method, microions can be introduced in a straightforward manner.
A recent comparative study \cite{Smiatek2009} indicated that the electrostatic interaction is the most expensive part in terms of the computational cost. 
Therefore, for intermediate or high salt concentrations, different methods for modelling the fluid becomes comparable.

For colloidal particles, one requirement of the simulation model is the realization of no-slip boundary condition on the colloid surface. 
In this work, we present such a colloid model based on the Dissipative Particle Dynamics, with full considerations of the hydrodynamic and electrostatic interactions.
We apply this model to study the dynamics of a charged colloidal particle under static electric fields.
The remainder of this article is organized as follows: 
in section \ref{sec:model}, we introduce the simulation model and describe relevant parameters for the system.   
We present the simulation results of electrophoretic mobility in section \ref{sec:result}.
Finally, we conclude in section \ref{sec:summary} with a brief summary.

\section{Simulation Model}
\label{sec:model}

Our simulation system consists of three parts: the solvent, the colloidal
particle and the microions.
All simulations were carried out using the open source package ESPResSo
\cite{ESPResSo}.

\subsection{Fluids}
\label{sec:fluid}

The fluids are simulated using Dissipative Particle Dynamics (DPD), an established method for mesoscale fluid simulations.
For two fluid beads $i$ and $j$, we denote their relative displacement as $\vec{r}_{ij}= \vec{r}_i - \vec{r}_j$, and their relative velocity $\vec{v}_{ij}= \vec{v}_i - \vec{v}_j$.
The distance between two beads is denoted by $r_{ij}=  |\vec{r}_{ij}|$ and the unit vector is $\hat{\vec{r}}_{ij}=  \vec{r}_{ij}/r_{ij}$.
The basic DPD equations involve the pair interaction between fluid beads.
The force exerted by bead $j$ on bead $i$ is given by
\begin{equation}
  \vec{F}_{ij}^{\rm DPD} = \vec{F}_{ij}^{\rm D} + \vec{F}_{ij}^{\rm R}.
\end{equation}
The dissipative part $\vec{F}_{ij}^{\rm D}$ is proportional to the
relative velocity between two fluid beads,
\begin{equation}  
  \label{eq:dpd_F_D}
  \vec{F}_{ij}^{\rm D} = - \gamma^{\rm DPD} \, \omega^{\rm D}(r_{ij}) 
  (\vec{v}_{ij} \cdot \hat{\vec{r}}_{ij}) 
  \hat{\vec{r}}_{ij},
\end{equation}
with a friction coefficient $\gamma^{\rm DPD}$.
The weight function $\omega^{\rm D}(r_{ij})$ is a monotonically decreasing function of $r_{ij}$, and vanishes at a given cutoff $r_{\rm c}^{\rm DPD}$,
\begin{equation}
  \omega^{\rm D} (r) = \begin{cases} 
    \left( 1 - \dfrac{r}{r_{\rm c}} \right)^2 & \quad \text{if } r \le r_{\rm c}^{\rm DPD},  \\
    0 & \quad \text{if } r>r_{\rm c}^{\rm DPD}. \end{cases}
\end{equation}
The cutoff radius $r_{\rm c}^{\rm DPD}$ characterizes the finite range of the interaction.

The random force $\vec{F}_{ij}^{\rm R}$ has the form
\begin{equation}
\vec{F}_{ij}^{\rm R} = \sqrt{ 2 k_B T \gamma^{\rm DPD} \,
\omega^{\rm D}(r_{ij}) } \, \xi_{ij} \hat{\vec{r}}_{ij},
\end{equation}
where $\xi_{ij}=\xi_{ji}$ are symmetric, but otherwise uncorrelated random
functions with zero mean and variance $\langle \xi_{ij}(t)
\xi_{ij}(t^{\prime}) \rangle = \delta(t-t^{\prime})$ (here $\delta(t)$ is
Dirac's delta function).
The fluctuation-dissipation theorem relates the magnitude of the stochastic contribution to the dissipative part, to ensure the correct equilibrium statistics.
The pair forces between two beads satisfy Newton's third law, $\vec{F}_{ij}=-\vec{F}_{ji}$, hence the momentum is conserved.
The momentum-conservation feature of DPD leads to the correct long-time hydrodynamic behavior (i.e. Navier-Stokes equations).

In the following, physical quantities will be reported in a model unit system of $\sigma$ (length), $m$ (mass), $\varepsilon$ (energy), $e$(charge) and a derived time unit $\tau=\sigma\sqrt{m/\varepsilon}$. 
We use the fluid density $\rho=3.0\,\sigma^{-3}$. 
The friction coefficient is $\gamma^{\rm DPD}=5.0\,m/\tau$ and the cutoff for DPD is $r_{\rm c}^{\rm DPD}=1.0\,\sigma$.
To measure the shear viscosity, we implement the method in Ref. \cite{Smiatek2008} to simulate the Poiseuille and Couette flows in a thin channel geometry. 
The viscosity of the fluid with our parameter setting is $\eta_s = 1.23 \pm 0.01 \, m/(\sigma\tau)$.

\subsection{Microions}
\label{sec:microion}

Microions (either counterions to balance the colloid charge or the dissolved electrolytes) are introduced as the same DPD beads as the fluid, but carry charges and have exclusive interactions (to other charged beads but not to the fluid beads). 
We only consider the monovalent case where microions carry a single elementary charge $\pm e$. 
The exclusive interaction is necessary to prevent the collapse of charged system. 
A modified, pure repulsive Lennard-Jones interaction is used \cite{WCA},
\begin{equation}
  \label{eq:WCA}
  V(r) = \begin{cases}
    4\varepsilon \left[ \left( \dfrac{\sigma}{r-r_0} \right)^{12} - 
      \left( \dfrac{\sigma}{r-r_0} \right)^6 + \dfrac{1}{4} \right] 
    & \text{if } r-r_0 \le r_{\rm c}^{\rm LJ}, \\ 
    0 &  \text{if } r-r_0 > r_{\rm c}^{\rm LJ},
  \end{cases}
\end{equation}
where $r$ is the distance between two charged beads. 
The cutoff radius is set at the potential minimum $r_{\rm c}^{\rm LJ}= \sqrt[6]{2}\, \sigma$. 
The microions have a size of $1.0\,\sigma$ ($r_0=0$). 
Charged microions also interact by Coulomb interactions, and we compute the electrostatic interactions using Particle-Particle-Particle Mesh (P3M) method \cite{HockneyEastwood, Deserno1998, Deserno1998a}.
The Bjerrum length $l_B=e^2/(4\pi \epsilon_m k_BT)$ of the fluid is set to $1.0\,\sigma$ and the temperature is $k_BT=1.0\,\varepsilon$. 

\begin{figure}[htbp]
  \centerline{\includegraphics[width=0.7\columnwidth]{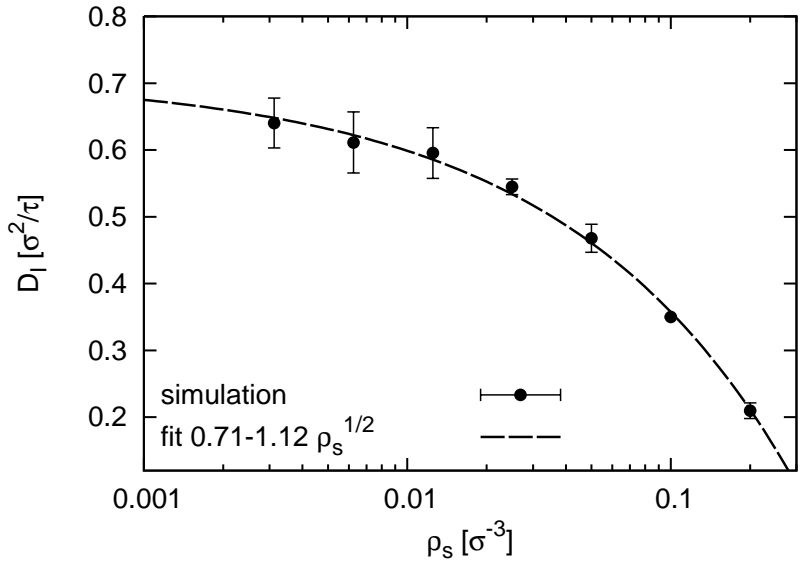}}
  \caption{The diffusion constant $D_{\rm I}$ of microions as a function of salt concentration $\rho_s$. Symbols are the simulation results. Error bars are obtained by averaging five independent simulation runs. The curve is a fit to Kohlrausch law with fitting parameters $A=0.71$ and $B=1.12$.}  
  \label{fig:DI}
\end{figure}

One useful quantity is the diffusion constant of the microion $D_{\rm I}$.
We measure the mean-square displacement of the microion, then use a linear regression at late times to obtain the diffusion constant, 
\begin{equation}
  \label{eq:Einstein}
  \lim_{ t \rightarrow \infty} \langle ( \vec{r}(t) - \vec{r}(0))^2 \rangle = 6 D_{\rm I} t .
\end{equation}
The diffusion constant depends on the salt concentration $\rho_s$. 
We perform simulations with different salt concentrations, $\rho_s=0.003125$ -- $0.2\,\sigma^{-3}$, in a simulation box $L=30\,\sigma$.    
The simulation results are compared with the empirical Kohlrausch law \cite{Wright}, which states that microion's diffusion constant depends linearly on the square root of the salt concentration $\sqrt{\rho_s}$,  
\begin{equation}
  D_{\rm I} = A - B \sqrt{\rho_s}, 
\end{equation}
where $A$ and $B$ are two fitting parameters. 
Fig.~\ref{fig:DI} shows the simulation results and a fit to Kohlrausch law.  

\subsection{Colloid}
\label{sec:colloid}

The colloidal particle is represented by a large sphere which has modified Lennard-Jones type conservative interaction to the fluid beads.
The interaction has a similar form of Eq. (\ref{eq:WCA}), but with a larger radius $R=r_0+\sigma=3.0\,\sigma$. 

\begin{figure}[htbp]
  \centerline{\includegraphics[width=0.7\columnwidth]{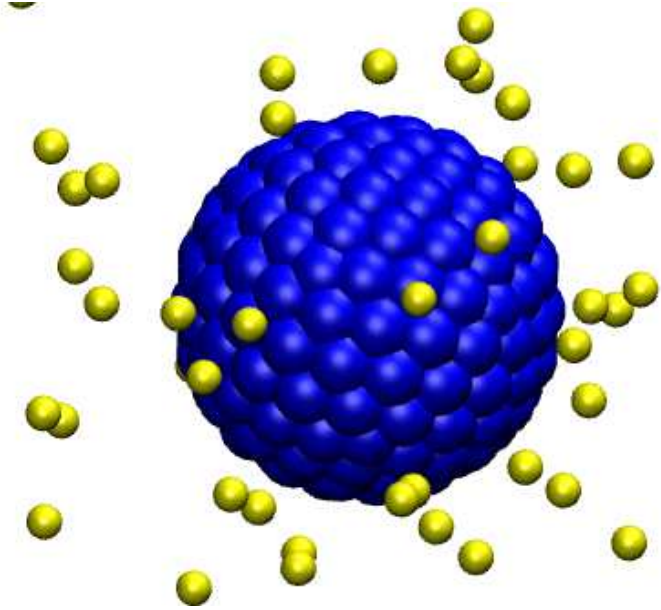}}
  \caption{Snapshot of a colloidal particle in a salt-free solution. The
surface sites are represented by the dark beads, and the light beads are
counterions. For clarity, solvent beads are not shown here. }
  \label{fig:snap}
\end{figure}

To implement the boundary condition at the surface, a set of DPD interaction sites is distributed evenly on the surface $R=3.0\,\sigma$, and the number of the sites is $N_s$. 
The position of the interacting sites is fixed with respect to the colloid center. 
These sites interact with the solvent beads through the DPD dissipative and stochastic interactions, with the friction constant $\gamma^{\rm CS}$ and the same cutoff $r_{\rm c}^{\rm DPD}$. 
The total force exerted on the colloid is given by the sum over all DPD
interactions due to the surface sites, plus the conservative excluded volume
interaction,
\begin{equation}
  \vec{F}^{\rm C} = \sum^{N_s}_{i=1} \vec{F}^{\rm DPD}_i(\vec{r}_i) + \vec{F}^{\rm LJ}.
\end{equation}
Here $\vec{r}_i$ denotes the position of $i$-th surface sites.
Similarly, the torque exerted on the colloid can be written as
\begin{equation}
  \vec{T}^{\rm C} = \sum^{N_s}_{i=1} \vec{F}^{\rm DPD}_i (\vec{r}_i)  \times ( \vec{r}_i - \vec{r}_{\rm cm}),
\end{equation}
where $\vec{r}_{\rm cm}$ is the position vector of the colloid's center-of-mass.
Note that the excluded volume interaction does not contribute to the torque because the associated force points towards the colloid center.  
The total force and torque are then used to update the position and velocity of the colloid in a time step using the Velocity-Verlet algorithm.  
The mass of the colloidal particle is $M=100\,m$ and the moment of inertia is $I=360\,m\sigma^2$, corresponding to a uniformly distributed mass. 
Fig.~\ref{fig:snap} shows a representative snapshot of a single charged colloidal particle with counterions.

\begin{figure}[htbp]
  \centerline{\includegraphics[width=0.7\columnwidth]{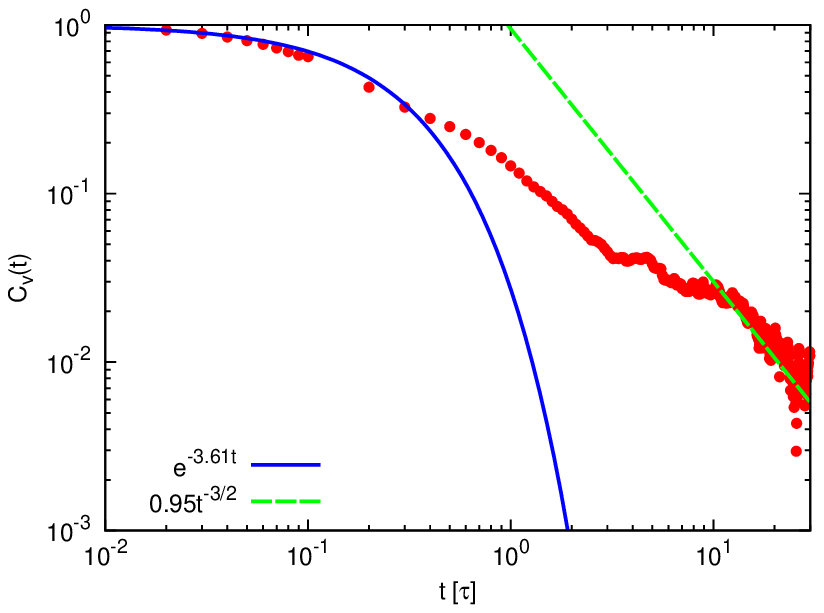}}
  \centerline{\includegraphics[width=0.7\columnwidth]{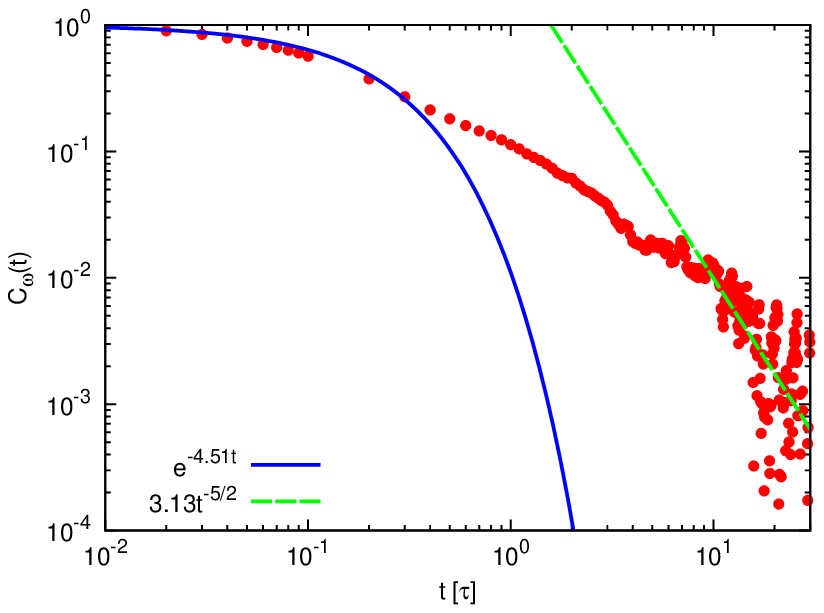}}
  \caption{Translational (top) and rotational (bottom) velocity autocorrelation
functions. The measurement is performed for a single uncharged colloid with radius
$R=3.0\sigma$ in a salt-free solution. The temperature is $k_BT = 1.0\, \varepsilon$ and surface-fluid DPD parameter $\gamma^{\rm CS}=10.0\, m/\tau$.} 
  \label{fig:acf}
\end{figure}

As a benchmark for our colloid model, we have performed simulations of an uncharged colloid in a simulation box $L=60\,\sigma$ and measured the autocorrelation functions.  
Two functions are obtained from the simulations: the translational and rotational velocity autocorrelation functions
\begin{eqnarray}
  \label{eq:vacf}
  C_v(t) &=& \frac{ \langle \vec{v}(0) \cdot \vec{v}(t) \rangle } 
  { \langle \vec{v}^2 \rangle }, \\
  \label{eq:wacf}
  C_{\omega}(t) &=& \frac{ \langle \vec{\omega}(0) \cdot \vec{\omega}(t) \rangle } 
  { \langle \vec{\omega}^2 \rangle }, 
\end{eqnarray}
where $\vec{v}(t)$ and $\vec{\omega}(t)$ are the translational velocity and rotational velocity of the colloid at time $t$, respectively. 
Fig.~\ref{fig:acf} shows the simulation results for $\gamma^{\rm CS}=10.0\,m/\tau$ in log-log plots.

For short time lags, both autocorrelation functions show exponential relaxation.
The decay rate can be calculated using the Enskog dense-gas kinetic theory
\cite{Subramanian1975, Hynes1977, Padding2005, Whitmer2010}
\begin{eqnarray}
  \label{eq:ens_v}
  \lim_{t \rightarrow 0} \, C_v(t) &=& \exp( -\zeta_{\rm ENS}^{v} t ), \\
  \label{eq:ens_w}
  \lim_{t \rightarrow 0} \, C_{\omega}(t) &=& \exp( -\zeta_{\rm ENS}^{\omega} t),
\end{eqnarray}
where the Enskog friction coefficients are 
\begin{eqnarray}
  \zeta_{\rm ENS}^{v}  &=& \frac{8}{3} \left( \frac{2\pi k_B T m M}{ m+M } \right)^{1/2} \rho R^2 \frac{2}{M} ,\\
    \zeta_{\rm ENS}^{\omega}  &=& \frac{8}{3} \left( \frac{2\pi k_B T m M}{ m+M } \right)^{1/2} \rho R^2 \frac{5}{2M},
\end{eqnarray}
where $m$ and $M$ are the mass for the fluid bead and the colloid, respectively, and $\rho$ is the solvent density.
Eqs. (\ref{eq:ens_v}) and (\ref{eq:ens_w}) are plotted as solid curves in Fig.~\ref{fig:acf} and show reasonable agreement with the simulation data when
$t<0.1\, \tau$.  

For long time lags, hydrodynamic effects set in and lead to a slow relaxation
for autocorrelation functions \cite{Alder1970}. 
This so-called long-time tail is the manifestation of momentum conservation, as the momentum must be transported away from the colloid in a diffusive manner.
Mode-coupling theory predicts an algebraic behavior at long times \cite{HansenMcDonald}
\begin{eqnarray}
  \lim_{t \rightarrow \infty} \, \langle \vec{v}(0) \cdot \vec{v}(t) \rangle  &=& \frac{k_BT} {12 m \rho [\pi (\nu+D) ]^{3/2} } \,\, t^{-3/2}, \\
  \lim_{t \rightarrow \infty} \, \langle \mbox{\boldmath${\omega}$}(0) \cdot \mbox{\boldmath${\omega}$}(t) \rangle &=& \frac{\pi k_BT } { m \rho [ 4\pi (\nu+D) ]^{5/2} } \,\, t^{-5/2},
\end{eqnarray}
where $\nu=\eta_s/\rho$ is the kinematic viscosity and $D$ is the diffusion constant of the colloid, which is much smaller than $\nu$ and can be neglected. 
The results are plotted as dashed lines in Fig.~\ref{fig:acf}. The data are consistent with the theoretical prediction for $t>10\,\tau$, but the rotational autocorrelation function exhibits large fluctuations for large times.  
This is mainly due to the fact that the statistics for long-time values becomes very bad, and very long simulations are required to obtain accurate values.

\begin{figure}[htbp]
  \centerline{\includegraphics[width=0.7\columnwidth]{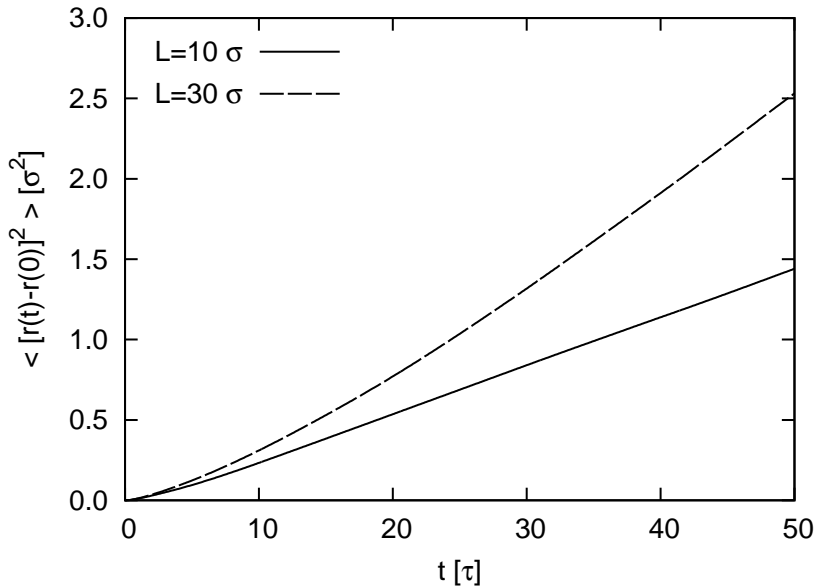}}
  \caption{The mean-square displacement of a spherical colloid with radius $R=3.0\, \sigma$ for two different sizes of simulation box, $L=10\, \sigma$ and $L=30\, \sigma$.}
  \label{fig:msd}
\end{figure}

The diffusion constant of the colloid can be calculated from the velocity autocorrelation function using the Green-Kubo relation
\begin{equation}
  D = \frac{1}{3} \int_0^{\infty} {\rm d}t \, \langle \vec{v}(0) \cdot \vec{v}(t) \rangle. 
\end{equation}
Alternatively, the diffusion constant can be obtained from the mean-square displacement, similar to Eq. (\ref{eq:Einstein}). 
Due to the periodic boundary condition implemented in simulations, the diffusion constant for a colloid in a finite simulation box depends on the box size. 
Fig.~\ref{fig:msd} demonstrates the finite-size effect by plotting the mean-square displacement as a function of time for two different simulation boxes $L=10\, \sigma$ and $L=30\, \sigma$. 

\begin{figure}[htbp]
  \centerline{\includegraphics[width=0.7\columnwidth]{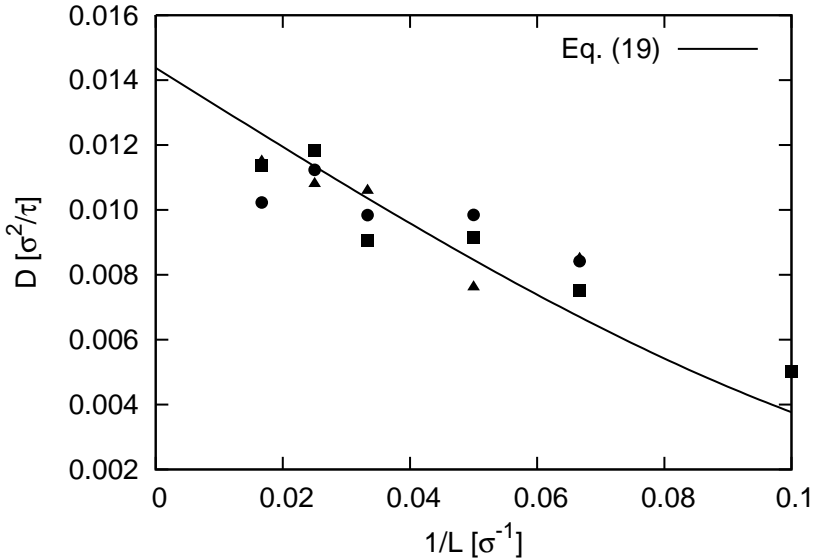}}
  \caption{The diffusion constant $D$ for a spherical colloid of radius $R=3.0 \, \sigma$ as a function of the reciprocal of the box size $1/L$. The curve is the prediction from Eq. (\ref{eq:Hasimoto}). Different symbols correspond to simulation runs with different initialization of the random generator.}
  \label{fig:box}
\end{figure}

The diffusion constant increases with increasing box size. 
For small simulation box, the long-wavelength hydrodynamic modes are suppressed due to the coupling between the colloid and its periodic images. 
An analytic expression for the diffusion constant in terms of an expansion of powers of $1/L$ was derived by Hasimoto \cite{Hasimoto1959}
\begin{equation}
  \label{eq:Hasimoto}
  D = \frac{ k_BT }{ 6\pi \eta_s } \left( \frac{1}{R} - \frac{2.837}{L} + \frac{4.19 R^2}{L^3} + \cdots \right).
\end{equation}
In Fig.~\ref{fig:box}, simulation results of the diffusion constant are plotted in terms of $1/L$, the reciprocal of the box size, and the curve is the prediction from Eq. (\ref{eq:Hasimoto}).
The simulation results and the hydrodynamic theory agree well.

\begin{figure}[htbp]
  \centerline{\includegraphics[width=0.7\columnwidth]{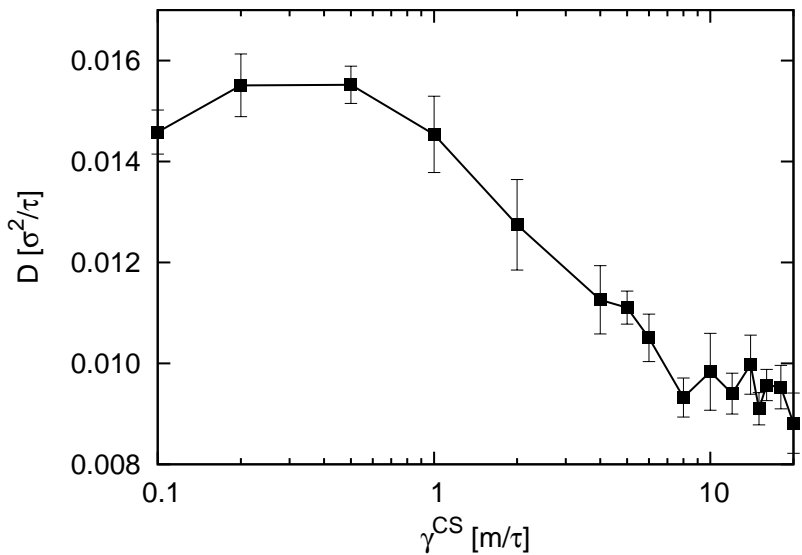}}
  \caption{The diffusion constant $D$ for a spherical colloid of radius $R=3.0 \, \sigma$ as a function of the surface-fluid DPD friction coefficient $\gamma^{\rm CS}$. The simulation box has a size of $L=30\,\sigma$.}
  \label{fig:gamma}
\end{figure}

Recently, studies of flow over superhydrophobic surfaces demonstrate that no-slip boundary condition is not always appropriate \cite{Bocquet2007, Vinogradova2011}.
A more general boundary condition is the Navier boundary condition, where finite slip over the surface is allowed.
One advantage of our colloid model is the ability to adjust the boundary condition from no-slip to full-slip by changing the surface-fluid DPD friction $\gamma^{\rm CS}$. 
Fig.~\ref{fig:gamma} illustrates the change of the diffusion constant by varying $\gamma^{\rm CS}$ in a simulation box $L=30\,\sigma$. 
This freedom provides opportunities to study the effect of hydrodynamic slip on the colloidal dynamics \cite{Swan2008, Khair2009}.

\section{Electrophoretic mobility}
\label{sec:result}

In this section, we apply our DPD-based colloid model to investigate the electrophoretic mobility of a charged colloid, and compare simulation results with predictions from electrokinetic theories \cite{RSS, OBrien1978, Hill2003a}.

Charged colloids in an aqueous solution are surrounded by counterions and dissolved salt ions. 
Counterions accumulate around the colloid surface due to the Coulomb attraction between opposite charges and form an electric double layer (EDL). 
In equilibrium, the counterion cloud has spherical symmetry for a spherical colloid.
The thickness of the EDL is characterized by the Debye screening length 
\begin{equation}
  l_D = \kappa^{-1} = \left[ 4\pi l_B \sum_i z_i^2 \rho_i (\infty) \right]^{ - \frac{1}{2} },
\end{equation}
where the summation runs over different ion species; $z_i$ and $\rho_i(\infty)$ are the valence and the bulk concentration for $i$-th ion, respectively.
When an external electric field is applied to the suspension, the colloid (assumed to be positively charged) starts to move in the direction of the field, while the counterion cloud (negatively charged) is deformed and elongated in the opposite direction of the field. 
A steady state is reached when the electric driving force is equal to the hydrodynamic friction acting on the colloid. 
When the field strength is small, the final velocity of the colloid $\vec{u}$ depends linearly on the applied electric filed 
$\vec{E}$, and the proportionality defines the electrophoretic mobility $\mu$:
\begin{equation}
  \vec{u} = \mu \vec{E}.
\end{equation}
The electrophoretic mobility is in general a second-order tensor, but is reduced to a scalar for spherical colloids. 
Due to the complexity of the system and the coupling between the hydrodynamic and electrostatic interactions, analytic solutions for the mobility only exist for limiting cases.

In the literature, the electrophoretic mobility is often expressed in terms of a $\zeta$-potential, defined as the electrostatic potential at the shear plane. 
In the limit of $\kappa R \ll 1$ where the thickness of electric double layer is much larger than the colloidal size, the mobility is given by the H\"uckel formula \cite{Hueckel1924}
\begin{equation}
  \label{eq:Hueckel}
  \mu_{\rm H} = \frac{ 2 \epsilon_m \zeta }{ 3 \eta_s },
\end{equation}
where $\epsilon_m$ is the fluid permittivity and $\eta_s$ is the shear viscosity. 
In the opposite limit when the Debye screen length is much thinner in comparison to the colloid size ($\kappa R \gg 1$), the famous Smoluchowski's formula states \cite{Smoluchowski1917}
\begin{equation}
  \label{eq:Smoluchowshi}
  \mu_{\rm S} = \frac{ \epsilon_m \zeta } { \eta_s } = \frac{3}{2} \mu_{\rm H}.
\end{equation}
For more general cases of intermediate values of $\kappa R$, one has to rely on numerical methods to solve the electrokinetic equations \cite{OBrien1978, Hill2003a}.
In this work, we compute the electrophoretic mobility using the software MPEK
\footnote{http://reghanhill.research.mcgill.ca/research/mpek.html}.

In simulations, we use the colloid charge as a controlling parameter. 
To convert the colloid charge to the zeta-potential, one need to solve the Poisson-Boltzmann equation. 
For a spherical particle, numerical tables for the solution to Poisson-Boltzmann equation were given by Loeb \emph{et al.} \cite{LOW_table}. 
An analytic expression for the relationship between the $\zeta$-potential and the surface charge density $\sigma$ was derived by Ohshima \emph{et al.} \cite{Ohshima1982, Ohshima}
\begin{eqnarray}
  \sigma = \frac{2\epsilon_m \kappa k_B T}{e} \sinh \left( \frac{e\zeta}{2k_BT} \right) && \bigg[ 1 + \frac{1}{\kappa R} \frac{2}{\cosh^2(e\zeta/4k_BT)} \nonumber \\
  && + \frac{1}{(\kappa R)^2} \frac{ 8\ln [ \cosh(e\zeta/4k_BT)] }{ \sinh^2(e\zeta/2k_BT) } \bigg]^{1/2} .
  \label{eq:Ohshima}
\end{eqnarray}
We use the above equation to compute the zeta-potential from known surface charge density.

\begin{figure}[htbp]
  \centerline{\includegraphics[width=0.7\columnwidth]{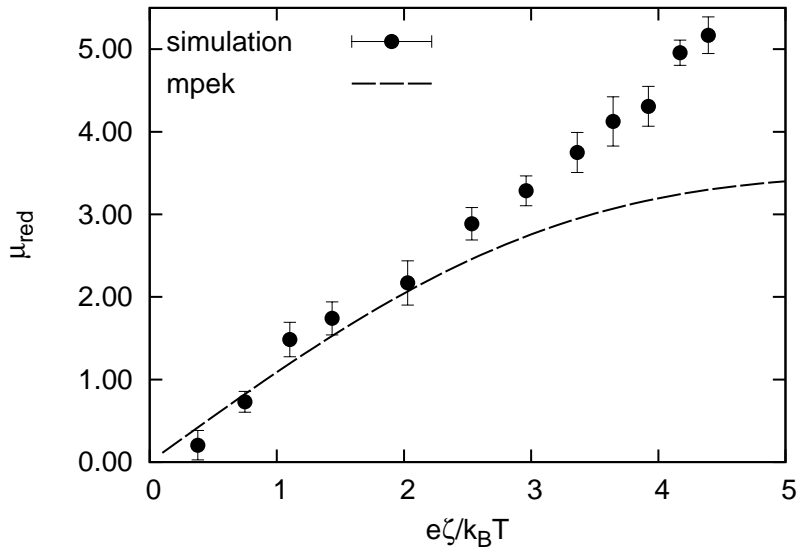}}
  \caption{The reduced mobility as a function of $\zeta$-potential for a spherical colloid of size $R=3.0\, \sigma$ and salt density $\rho_s=0.05\, \sigma^{-3}$.}
  \label{fig:mu_zeta}
\end{figure}

Fig.~\ref{fig:mu_zeta} plots the electrophoretic mobility as a function of the dimensionless zeta-potential $e\zeta/k_BT$. 
The simulations are performed for a charged spherical colloid of radius $R=3.0\,\sigma$, and the solution has a salt concentration $\rho_s=0.05 \, \sigma^{-3}$. 
We follow the standard to use the reduced mobility, which is a dimensionless number and is defined as
\begin{equation}
  \mu_{\rm red} = \frac{ 6 \pi \eta_s l_B } { e } \mu.
\end{equation}
The simulation results agree well with the prediction from the electrokinetic theory at small zeta potentials. 
At large zeta potential, the steric effects of the microions may play a role \cite{Lopez-Garcia2007, Lopez-Garcia2008, Khair2009}, resulting in an increase of the mobility.

\begin{figure}[htbp]
  \centerline{\includegraphics[width=0.7\columnwidth]{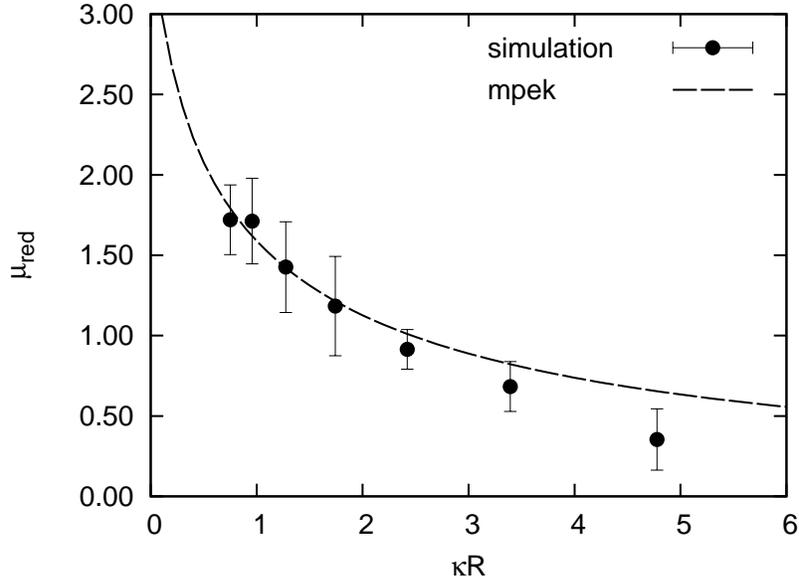}}
  \caption{The reduced mobility as a function of $\kappa R$ for a spherical colloid of size $R=3.0\, \sigma$ and charge $Q=10\, e$.}
  \label{fig:mu_ka}
\end{figure}

Fig.~\ref{fig:mu_ka} plots the reduced electrophoretic mobility as a function of $\kappa R$.
In simulations, the colloid has a fixed radius $R=3.0\,\sigma$, and the salt concentration is varied to change the value of $\kappa R$.
The charge on the colloid is small, $Q=10\, e$, to ensure that the zeta-potential is also small. 
The simulation results and the theoretic predictions agree well, except at very large salt concentration (large $\kappa R$ value). 
One possible explanation to the discrepancy is the change of microion's diffusion constant with respect to the salt concentration (see Fig.~\ref{fig:DI}), which is not taken into account in the electrokinetic theory. 

\section{Summary}
\label{sec:summary}

We have developed a mesoscopic colloid model based on the Dissipative Particle Dynamics. 
We have taken accounts in full for the hydrodynamic interaction with thermal fluctuations, using Dissipative Particle Dynamics, and the electrostatic interactions, using Particle-Particle-Particle Mesh method. 
We applied this new colloid model to investigate the electrophoretic mobility of a charged colloid under a static external field.
The simulation results show good agreement with the predictions from electrokinetic theories.
Futhermore, this model has been applied successfully to study the dynamic and dielectric response of a charged colloid to alternating electric fields \cite{2012_ac, 2013_q0, 2013_response}.

\begin{acknowledgments}
We are grateful to Prof. Reghan Hill for providing the computer program MPEK. 
We thank the HLRS Stuttgart for a generous grant of computer time on HERMIT.
This work is funded by the Deutsche Forschungsgemeinschaft (DFG) through the SFB-TR6 program ``Physics of Colloidal Dispersions in External Fields''.  
\end{acknowledgments}

\bibliography{CCAC}

\begin{thebibliography}{53}
\expandafter\ifx\csname natexlab\endcsname\relax\def\natexlab#1{#1}\fi
\expandafter\ifx\csname bibnamefont\endcsname\relax
  \def\bibnamefont#1{#1}\fi
\expandafter\ifx\csname bibfnamefont\endcsname\relax
  \def\bibfnamefont#1{#1}\fi
\expandafter\ifx\csname citenamefont\endcsname\relax
  \def\citenamefont#1{#1}\fi
\expandafter\ifx\csname url\endcsname\relax
  \def\url#1{\texttt{#1}}\fi
\expandafter\ifx\csname urlprefix\endcsname\relax\def\urlprefix{URL }\fi
\providecommand{\bibinfo}[2]{#2}
\providecommand{\eprint}[2][]{\url{#2}}

\bibitem[{\citenamefont{Russel et~al.}(1989)\citenamefont{Russel, Saville, and
  Schowalter}}]{RSS}
\bibinfo{author}{\bibfnamefont{W.~B.} \bibnamefont{Russel}},
  \bibinfo{author}{\bibfnamefont{D.~A.} \bibnamefont{Saville}},
  \bibnamefont{and}
  \bibinfo{author}{\bibfnamefont{W.}~\bibnamefont{Schowalter}},
  \emph{\bibinfo{title}{Colloidal Dispersions}} (\bibinfo{publisher}{Cambridge
  University Press}, \bibinfo{address}{Cambridge}, \bibinfo{year}{1989}).

\bibitem[{\citenamefont{Dhont}(1996)}]{Dhont}
\bibinfo{author}{\bibfnamefont{J.}~\bibnamefont{Dhont}},
  \emph{\bibinfo{title}{An Introduction to Dynamics of Colloids}}
  (\bibinfo{publisher}{Elsevier}, \bibinfo{address}{Amsterdam},
  \bibinfo{year}{1996}).

\bibitem[{\citenamefont{Ahlrichs and D\"unweg}(1999)}]{Ahlrichs1999}
\bibinfo{author}{\bibfnamefont{P.}~\bibnamefont{Ahlrichs}} \bibnamefont{and}
  \bibinfo{author}{\bibfnamefont{B.}~\bibnamefont{D\"unweg}},
  \bibinfo{journal}{J. Chem. Phys.} \textbf{\bibinfo{volume}{111}},
  \bibinfo{pages}{8225} (\bibinfo{year}{1999}).

\bibitem[{\citenamefont{Grass et~al.}(2008)\citenamefont{Grass, B\"ohme,
  Scheler, Cottet, and Holm}}]{Grass2008}
\bibinfo{author}{\bibfnamefont{K.}~\bibnamefont{Grass}},
  \bibinfo{author}{\bibfnamefont{U.}~\bibnamefont{B\"ohme}},
  \bibinfo{author}{\bibfnamefont{U.}~\bibnamefont{Scheler}},
  \bibinfo{author}{\bibfnamefont{H.}~\bibnamefont{Cottet}}, \bibnamefont{and}
  \bibinfo{author}{\bibfnamefont{C.}~\bibnamefont{Holm}},
  \bibinfo{journal}{Phys. Rev. Lett.} \textbf{\bibinfo{volume}{100}},
  \bibinfo{pages}{096104} (\bibinfo{year}{2008}).

\bibitem[{\citenamefont{Grass and Holm}(2009)}]{Grass2009}
\bibinfo{author}{\bibfnamefont{K.}~\bibnamefont{Grass}} \bibnamefont{and}
  \bibinfo{author}{\bibfnamefont{C.}~\bibnamefont{Holm}},
  \bibinfo{journal}{Soft Matter} \textbf{\bibinfo{volume}{5}},
  \bibinfo{pages}{2079} (\bibinfo{year}{2009}).

\bibitem[{\citenamefont{Grass and Holm}(2010)}]{Grass2010}
\bibinfo{author}{\bibfnamefont{K.}~\bibnamefont{Grass}} \bibnamefont{and}
  \bibinfo{author}{\bibfnamefont{C.}~\bibnamefont{Holm}},
  \bibinfo{journal}{Faraday Discuss.} \textbf{\bibinfo{volume}{144}},
  \bibinfo{pages}{57} (\bibinfo{year}{2010}).

\bibitem[{\citenamefont{Lobaskin and D\"unweg}(2004)}]{Lobaskin2004}
\bibinfo{author}{\bibfnamefont{V.}~\bibnamefont{Lobaskin}} \bibnamefont{and}
  \bibinfo{author}{\bibfnamefont{B.}~\bibnamefont{D\"unweg}},
  \bibinfo{journal}{New Journal of Physics} \textbf{\bibinfo{volume}{6}},
  \bibinfo{pages}{54} (\bibinfo{year}{2004}).

\bibitem[{\citenamefont{Lobaskin et~al.}(2004)\citenamefont{Lobaskin, D\"unweg,
  and Holm}}]{Lobaskin2004a}
\bibinfo{author}{\bibfnamefont{V.}~\bibnamefont{Lobaskin}},
  \bibinfo{author}{\bibfnamefont{B.}~\bibnamefont{D\"unweg}}, \bibnamefont{and}
  \bibinfo{author}{\bibfnamefont{C.}~\bibnamefont{Holm}}, \bibinfo{journal}{J.
  Phys.: Condens. Matter} \textbf{\bibinfo{volume}{16}}, \bibinfo{pages}{S4063}
  (\bibinfo{year}{2004}).

\bibitem[{\citenamefont{Lobaskin et~al.}(2007)\citenamefont{Lobaskin, D\"unweg,
  Medebach, Palberg, and Holm}}]{Lobaskin2007}
\bibinfo{author}{\bibfnamefont{V.}~\bibnamefont{Lobaskin}},
  \bibinfo{author}{\bibfnamefont{B.}~\bibnamefont{D\"unweg}},
  \bibinfo{author}{\bibfnamefont{M.}~\bibnamefont{Medebach}},
  \bibinfo{author}{\bibfnamefont{T.}~\bibnamefont{Palberg}}, \bibnamefont{and}
  \bibinfo{author}{\bibfnamefont{C.}~\bibnamefont{Holm}},
  \bibinfo{journal}{Phys. Rev. Lett.} \textbf{\bibinfo{volume}{98}},
  \bibinfo{pages}{176105} (\bibinfo{year}{2007}).

\bibitem[{\citenamefont{Chatterji and Horbach}(2005)}]{Chatterji2005}
\bibinfo{author}{\bibfnamefont{A.}~\bibnamefont{Chatterji}} \bibnamefont{and}
  \bibinfo{author}{\bibfnamefont{J.}~\bibnamefont{Horbach}},
  \bibinfo{journal}{J. Chem. Phys.} \textbf{\bibinfo{volume}{122}},
  \bibinfo{pages}{184903} (\bibinfo{year}{2005}).

\bibitem[{\citenamefont{Chatterji and Horbach}(2007)}]{Chatterji2007}
\bibinfo{author}{\bibfnamefont{A.}~\bibnamefont{Chatterji}} \bibnamefont{and}
  \bibinfo{author}{\bibfnamefont{J.}~\bibnamefont{Horbach}},
  \bibinfo{journal}{J. Chem. Phys.} \textbf{\bibinfo{volume}{126}},
  \bibinfo{pages}{064907} (\bibinfo{year}{2007}).

\bibitem[{\citenamefont{Giupponi and Pagonabarraga}(2011)}]{Giupponi2011}
\bibinfo{author}{\bibfnamefont{G.}~\bibnamefont{Giupponi}} \bibnamefont{and}
  \bibinfo{author}{\bibfnamefont{I.}~\bibnamefont{Pagonabarraga}},
  \bibinfo{journal}{Phys. Rev. Lett.} \textbf{\bibinfo{volume}{106}},
  \bibinfo{pages}{248304} (\bibinfo{year}{2011}).

\bibitem[{\citenamefont{Tanaka and Araki}(2000)}]{Tanaka2000}
\bibinfo{author}{\bibfnamefont{H.}~\bibnamefont{Tanaka}} \bibnamefont{and}
  \bibinfo{author}{\bibfnamefont{T.}~\bibnamefont{Araki}},
  \bibinfo{journal}{Phys. Rev. Lett.} \textbf{\bibinfo{volume}{85}},
  \bibinfo{pages}{1338} (\bibinfo{year}{2000}).

\bibitem[{\citenamefont{Nakayama and Yamamoto}(2005)}]{Nakayama2005}
\bibinfo{author}{\bibfnamefont{Y.}~\bibnamefont{Nakayama}} \bibnamefont{and}
  \bibinfo{author}{\bibfnamefont{R.}~\bibnamefont{Yamamoto}},
  \bibinfo{journal}{Phys. Rev. E} \textbf{\bibinfo{volume}{71}},
  \bibinfo{pages}{036707} (\bibinfo{year}{2005}).

\bibitem[{\citenamefont{Kim et~al.}(2006)\citenamefont{Kim, Nakayama, and
  Yamamoto}}]{Kim2006}
\bibinfo{author}{\bibfnamefont{K.}~\bibnamefont{Kim}},
  \bibinfo{author}{\bibfnamefont{Y.}~\bibnamefont{Nakayama}}, \bibnamefont{and}
  \bibinfo{author}{\bibfnamefont{R.}~\bibnamefont{Yamamoto}},
  \bibinfo{journal}{Phys. Rev. Lett.} \textbf{\bibinfo{volume}{96}},
  \bibinfo{pages}{208302} (\bibinfo{year}{2006}).

\bibitem[{\citenamefont{Nakayama et~al.}(2008)\citenamefont{Nakayama, Kim, and
  Yamamoto}}]{Nakayama2008}
\bibinfo{author}{\bibfnamefont{Y.}~\bibnamefont{Nakayama}},
  \bibinfo{author}{\bibfnamefont{K.}~\bibnamefont{Kim}}, \bibnamefont{and}
  \bibinfo{author}{\bibfnamefont{R.}~\bibnamefont{Yamamoto}},
  \bibinfo{journal}{Eur. Phys. J. E} \textbf{\bibinfo{volume}{26}},
  \bibinfo{pages}{361} (\bibinfo{year}{2008}).

\bibitem[{\citenamefont{Malevanets and Kapral}(1999)}]{Malevanets1999}
\bibinfo{author}{\bibfnamefont{A.}~\bibnamefont{Malevanets}} \bibnamefont{and}
  \bibinfo{author}{\bibfnamefont{R.}~\bibnamefont{Kapral}},
  \bibinfo{journal}{J. Chem. Phys.} \textbf{\bibinfo{volume}{110}},
  \bibinfo{pages}{8605} (\bibinfo{year}{1999}).

\bibitem[{\citenamefont{Gompper et~al.}(2009)\citenamefont{Gompper, Ihle,
  Kroll, and Winkler}}]{Gompper2009}
\bibinfo{author}{\bibfnamefont{G.}~\bibnamefont{Gompper}},
  \bibinfo{author}{\bibfnamefont{T.}~\bibnamefont{Ihle}},
  \bibinfo{author}{\bibfnamefont{D.~M.} \bibnamefont{Kroll}}, \bibnamefont{and}
  \bibinfo{author}{\bibfnamefont{R.~G.} \bibnamefont{Winkler}},
  \bibinfo{journal}{Adv. Polym. Sci.} \textbf{\bibinfo{volume}{221}},
  \bibinfo{pages}{1} (\bibinfo{year}{2009}).

\bibitem[{\citenamefont{Hoogerbrugge and Koelman}(1992)}]{Hoogerbrugge1992}
\bibinfo{author}{\bibfnamefont{P.~J.} \bibnamefont{Hoogerbrugge}}
  \bibnamefont{and} \bibinfo{author}{\bibfnamefont{J.~M. V.~A.}
  \bibnamefont{Koelman}}, \bibinfo{journal}{Europhys. Lett.}
  \textbf{\bibinfo{volume}{19}}, \bibinfo{pages}{155} (\bibinfo{year}{1992}).

\bibitem[{\citenamefont{Koelman and Hoogerbrugge}(1993)}]{Koelman1993}
\bibinfo{author}{\bibfnamefont{J.~M. V.~A.} \bibnamefont{Koelman}}
  \bibnamefont{and} \bibinfo{author}{\bibfnamefont{P.~J.}
  \bibnamefont{Hoogerbrugge}}, \bibinfo{journal}{Europhys. Lett.}
  \textbf{\bibinfo{volume}{21}}, \bibinfo{pages}{363} (\bibinfo{year}{1993}).

\bibitem[{\citenamefont{Espa\~nol and Warren}(1995)}]{Espanol1995}
\bibinfo{author}{\bibfnamefont{P.}~\bibnamefont{Espa\~nol}} \bibnamefont{and}
  \bibinfo{author}{\bibfnamefont{P.~B.} \bibnamefont{Warren}},
  \bibinfo{journal}{Europhys. Lett.} \textbf{\bibinfo{volume}{30}},
  \bibinfo{pages}{191} (\bibinfo{year}{1995}).

\bibitem[{\citenamefont{Groot and Warren}(1997)}]{Groot1997}
\bibinfo{author}{\bibfnamefont{R.~D.} \bibnamefont{Groot}} \bibnamefont{and}
  \bibinfo{author}{\bibfnamefont{P.~B.} \bibnamefont{Warren}},
  \bibinfo{journal}{J. Chem. Phys.} \textbf{\bibinfo{volume}{107}},
  \bibinfo{pages}{4423} (\bibinfo{year}{1997}).

\bibitem[{\citenamefont{Smiatek et~al.}(2009)\citenamefont{Smiatek, Sega, Holm,
  Schiller, and Schmid}}]{Smiatek2009}
\bibinfo{author}{\bibfnamefont{J.}~\bibnamefont{Smiatek}},
  \bibinfo{author}{\bibfnamefont{M.}~\bibnamefont{Sega}},
  \bibinfo{author}{\bibfnamefont{C.}~\bibnamefont{Holm}},
  \bibinfo{author}{\bibfnamefont{U.~D.} \bibnamefont{Schiller}},
  \bibnamefont{and} \bibinfo{author}{\bibfnamefont{F.}~\bibnamefont{Schmid}},
  \bibinfo{journal}{J. Chem. Phys.} \textbf{\bibinfo{volume}{130}},
  \bibinfo{pages}{244702} (\bibinfo{year}{2009}).

\bibitem[{\citenamefont{Limbach et~al.}(2006)\citenamefont{Limbach, Arnold,
  Mann, and Holm}}]{ESPResSo}
\bibinfo{author}{\bibfnamefont{H.}~\bibnamefont{Limbach}},
  \bibinfo{author}{\bibfnamefont{A.}~\bibnamefont{Arnold}},
  \bibinfo{author}{\bibfnamefont{B.}~\bibnamefont{Mann}}, \bibnamefont{and}
  \bibinfo{author}{\bibfnamefont{C.}~\bibnamefont{Holm}},
  \bibinfo{journal}{Comp. Phys. Comm.} \textbf{\bibinfo{volume}{174}},
  \bibinfo{pages}{704} (\bibinfo{year}{2006}).

\bibitem[{\citenamefont{Smiatek et~al.}(2008)\citenamefont{Smiatek, Allen, and
  Schmid}}]{Smiatek2008}
\bibinfo{author}{\bibfnamefont{J.}~\bibnamefont{Smiatek}},
  \bibinfo{author}{\bibfnamefont{M.}~\bibnamefont{Allen}}, \bibnamefont{and}
  \bibinfo{author}{\bibfnamefont{F.}~\bibnamefont{Schmid}},
  \bibinfo{journal}{Eur. Phys. J. E} \textbf{\bibinfo{volume}{26}},
  \bibinfo{pages}{115} (\bibinfo{year}{2008}).

\bibitem[{\citenamefont{Weeks et~al.}(1971)\citenamefont{Weeks, Chandler, and
  Andersen}}]{WCA}
\bibinfo{author}{\bibfnamefont{J.~D.} \bibnamefont{Weeks}},
  \bibinfo{author}{\bibfnamefont{D.}~\bibnamefont{Chandler}}, \bibnamefont{and}
  \bibinfo{author}{\bibfnamefont{H.~C.} \bibnamefont{Andersen}},
  \bibinfo{journal}{J. Chem. Phys.} \textbf{\bibinfo{volume}{54}},
  \bibinfo{pages}{5237} (\bibinfo{year}{1971}).

\bibitem[{\citenamefont{Hockney and Eastwood}(1988)}]{HockneyEastwood}
\bibinfo{author}{\bibfnamefont{R.}~\bibnamefont{Hockney}} \bibnamefont{and}
  \bibinfo{author}{\bibfnamefont{J.}~\bibnamefont{Eastwood}},
  \emph{\bibinfo{title}{Computer Simulation Using Particles}}
  (\bibinfo{publisher}{Adam Hilger}, \bibinfo{address}{Bristol},
  \bibinfo{year}{1988}).

\bibitem[{\citenamefont{Deserno and Holm}(1998{\natexlab{a}})}]{Deserno1998}
\bibinfo{author}{\bibfnamefont{M.}~\bibnamefont{Deserno}} \bibnamefont{and}
  \bibinfo{author}{\bibfnamefont{C.}~\bibnamefont{Holm}}, \bibinfo{journal}{J.
  Chem. Phys.} \textbf{\bibinfo{volume}{109}}, \bibinfo{pages}{7678}
  (\bibinfo{year}{1998}{\natexlab{a}}).

\bibitem[{\citenamefont{Deserno and Holm}(1998{\natexlab{b}})}]{Deserno1998a}
\bibinfo{author}{\bibfnamefont{M.}~\bibnamefont{Deserno}} \bibnamefont{and}
  \bibinfo{author}{\bibfnamefont{C.}~\bibnamefont{Holm}}, \bibinfo{journal}{J.
  Chem. Phys.} \textbf{\bibinfo{volume}{109}}, \bibinfo{pages}{7694}
  (\bibinfo{year}{1998}{\natexlab{b}}).

\bibitem[{\citenamefont{Wright}(2007)}]{Wright}
\bibinfo{author}{\bibfnamefont{M.~R.} \bibnamefont{Wright}},
  \emph{\bibinfo{title}{An Introduction to Aqueous Electrolyte Solutions}}
  (\bibinfo{publisher}{Wiley}, \bibinfo{address}{Chichester},
  \bibinfo{year}{2007}).

\bibitem[{\citenamefont{Subramanian and Davis}(1975)}]{Subramanian1975}
\bibinfo{author}{\bibfnamefont{G.}~\bibnamefont{Subramanian}} \bibnamefont{and}
  \bibinfo{author}{\bibfnamefont{H.}~\bibnamefont{Davis}},
  \bibinfo{journal}{Phys. Rev. A} \textbf{\bibinfo{volume}{11}},
  \bibinfo{pages}{1430} (\bibinfo{year}{1975}).

\bibitem[{\citenamefont{Hynes}(1977)}]{Hynes1977}
\bibinfo{author}{\bibfnamefont{J.}~\bibnamefont{Hynes}},
  \bibinfo{journal}{Annu. Rev. Phys. Chem.} \textbf{\bibinfo{volume}{28}},
  \bibinfo{pages}{301} (\bibinfo{year}{1977}).

\bibitem[{\citenamefont{Padding et~al.}(2005)\citenamefont{Padding, Wysocki,
  L\"owen, and Louis}}]{Padding2005}
\bibinfo{author}{\bibfnamefont{J.~T.} \bibnamefont{Padding}},
  \bibinfo{author}{\bibfnamefont{A.}~\bibnamefont{Wysocki}},
  \bibinfo{author}{\bibfnamefont{H.}~\bibnamefont{L\"owen}}, \bibnamefont{and}
  \bibinfo{author}{\bibfnamefont{A.~A.} \bibnamefont{Louis}},
  \bibinfo{journal}{J. Phys.: Condens. Matter} \textbf{\bibinfo{volume}{17}},
  \bibinfo{pages}{S3393} (\bibinfo{year}{2005}).

\bibitem[{\citenamefont{Whitmer and Luijten}(2010)}]{Whitmer2010}
\bibinfo{author}{\bibfnamefont{J.~K.} \bibnamefont{Whitmer}} \bibnamefont{and}
  \bibinfo{author}{\bibfnamefont{E.}~\bibnamefont{Luijten}},
  \bibinfo{journal}{J. Phys.: Condens. Matter} \textbf{\bibinfo{volume}{22}},
  \bibinfo{pages}{104106} (\bibinfo{year}{2010}).

\bibitem[{\citenamefont{Alder and Wainwright}(1970)}]{Alder1970}
\bibinfo{author}{\bibfnamefont{B.~J.} \bibnamefont{Alder}} \bibnamefont{and}
  \bibinfo{author}{\bibfnamefont{T.~E.} \bibnamefont{Wainwright}},
  \bibinfo{journal}{Phys. Rev. A} \textbf{\bibinfo{volume}{1}},
  \bibinfo{pages}{18} (\bibinfo{year}{1970}).

\bibitem[{\citenamefont{Hansen and McDonald}(2006)}]{HansenMcDonald}
\bibinfo{author}{\bibfnamefont{J.-P.} \bibnamefont{Hansen}} \bibnamefont{and}
  \bibinfo{author}{\bibfnamefont{I.~R.} \bibnamefont{McDonald}},
  \emph{\bibinfo{title}{Theory of Simple Liquids}}
  (\bibinfo{publisher}{Academic Press}, \bibinfo{address}{London},
  \bibinfo{year}{2006}), \bibinfo{edition}{3rd} ed.

\bibitem[{\citenamefont{Hasimoto}(1959)}]{Hasimoto1959}
\bibinfo{author}{\bibfnamefont{H.}~\bibnamefont{Hasimoto}},
  \bibinfo{journal}{J. Fluid Mech.} \textbf{\bibinfo{volume}{5}},
  \bibinfo{pages}{317} (\bibinfo{year}{1959}).

\bibitem[{\citenamefont{Bocquet and Barrat}(2007)}]{Bocquet2007}
\bibinfo{author}{\bibfnamefont{L.}~\bibnamefont{Bocquet}} \bibnamefont{and}
  \bibinfo{author}{\bibfnamefont{J.-L.} \bibnamefont{Barrat}},
  \bibinfo{journal}{Soft Matter} \textbf{\bibinfo{volume}{3}},
  \bibinfo{pages}{685} (\bibinfo{year}{2007}).

\bibitem[{\citenamefont{Vinogradova and Belyaev}(2011)}]{Vinogradova2011}
\bibinfo{author}{\bibfnamefont{O.~I.} \bibnamefont{Vinogradova}}
  \bibnamefont{and} \bibinfo{author}{\bibfnamefont{A.~V.}
  \bibnamefont{Belyaev}}, \bibinfo{journal}{J. Phys.: Condens. Matter}
  \textbf{\bibinfo{volume}{23}}, \bibinfo{pages}{184104}
  (\bibinfo{year}{2011}).

\bibitem[{\citenamefont{Swan and Khair}(2008)}]{Swan2008}
\bibinfo{author}{\bibfnamefont{J.~W.} \bibnamefont{Swan}} \bibnamefont{and}
  \bibinfo{author}{\bibfnamefont{A.~S.} \bibnamefont{Khair}},
  \bibinfo{journal}{J. Fluid Mech.} \textbf{\bibinfo{volume}{606}},
  \bibinfo{pages}{115} (\bibinfo{year}{2008}).

\bibitem[{\citenamefont{Khair and Squires}(2009)}]{Khair2009}
\bibinfo{author}{\bibfnamefont{A.~S.} \bibnamefont{Khair}} \bibnamefont{and}
  \bibinfo{author}{\bibfnamefont{T.~M.} \bibnamefont{Squires}},
  \bibinfo{journal}{Phys. Fluids} \textbf{\bibinfo{volume}{21}},
  \bibinfo{pages}{042001} (\bibinfo{year}{2009}).

\bibitem[{\citenamefont{O'Brien and White}(1978)}]{OBrien1978}
\bibinfo{author}{\bibfnamefont{R.~W.} \bibnamefont{O'Brien}} \bibnamefont{and}
  \bibinfo{author}{\bibfnamefont{L.~R.} \bibnamefont{White}},
  \bibinfo{journal}{J. Chem. Soc.{,} Faraday Trans. 2}
  \textbf{\bibinfo{volume}{74}}, \bibinfo{pages}{1607} (\bibinfo{year}{1978}).

\bibitem[{\citenamefont{Hill et~al.}(2003)\citenamefont{Hill, Saville, and
  Russel}}]{Hill2003a}
\bibinfo{author}{\bibfnamefont{R.~J.} \bibnamefont{Hill}},
  \bibinfo{author}{\bibfnamefont{D.~A.} \bibnamefont{Saville}},
  \bibnamefont{and} \bibinfo{author}{\bibfnamefont{W.~B.}
  \bibnamefont{Russel}}, \bibinfo{journal}{J. Colloid Interface Sci.}
  \textbf{\bibinfo{volume}{258}}, \bibinfo{pages}{56} (\bibinfo{year}{2003}).

\bibitem[{\citenamefont{H\"uckel}(1924)}]{Hueckel1924}
\bibinfo{author}{\bibfnamefont{E.}~\bibnamefont{H\"uckel}},
  \bibinfo{journal}{Phys. Z.} \textbf{\bibinfo{volume}{25}},
  \bibinfo{pages}{204} (\bibinfo{year}{1924}).

\bibitem[{\citenamefont{Smoluchowski}(1917)}]{Smoluchowski1917}
\bibinfo{author}{\bibfnamefont{M.~v.} \bibnamefont{Smoluchowski}},
  \bibinfo{journal}{Z. Phys. Chem.} \textbf{\bibinfo{volume}{92}},
  \bibinfo{pages}{129} (\bibinfo{year}{1917}).

\bibitem[{\citenamefont{Loeb et~al.}(1961)\citenamefont{Loeb, Overbeek, and
  Wiersema}}]{LOW_table}
\bibinfo{author}{\bibfnamefont{A.~L.} \bibnamefont{Loeb}},
  \bibinfo{author}{\bibfnamefont{J.~T.~G.} \bibnamefont{Overbeek}},
  \bibnamefont{and} \bibinfo{author}{\bibfnamefont{P.~H.}
  \bibnamefont{Wiersema}}, \emph{\bibinfo{title}{The Electrical Double Layer
  around a Spherical Colloid Particle}} (\bibinfo{publisher}{MIT Press},
  \bibinfo{address}{Massachusetts}, \bibinfo{year}{1961}).

\bibitem[{\citenamefont{Ohshima et~al.}(1982)\citenamefont{Ohshima, Healy, and
  White}}]{Ohshima1982}
\bibinfo{author}{\bibfnamefont{H.}~\bibnamefont{Ohshima}},
  \bibinfo{author}{\bibfnamefont{T.}~\bibnamefont{Healy}}, \bibnamefont{and}
  \bibinfo{author}{\bibfnamefont{L.}~\bibnamefont{White}}, \bibinfo{journal}{J.
  Colloid Interface Sci.} \textbf{\bibinfo{volume}{90}}, \bibinfo{pages}{17}
  (\bibinfo{year}{1982}).

\bibitem[{\citenamefont{Ohshima}(2006)}]{Ohshima}
\bibinfo{author}{\bibfnamefont{H.}~\bibnamefont{Ohshima}},
  \emph{\bibinfo{title}{Theory of Colloid and Interfacial Electric Phenomena}}
  (\bibinfo{publisher}{Academic Press}, \bibinfo{address}{Amsterdam},
  \bibinfo{year}{2006}).

\bibitem[{\citenamefont{L\'opez-Garc\'ia
  et~al.}(2007)\citenamefont{L\'opez-Garc\'ia, Aranda-Rasc\'on, and
  Horno}}]{Lopez-Garcia2007}
\bibinfo{author}{\bibfnamefont{J.}~\bibnamefont{L\'opez-Garc\'ia}},
  \bibinfo{author}{\bibfnamefont{M.}~\bibnamefont{Aranda-Rasc\'on}},
  \bibnamefont{and} \bibinfo{author}{\bibfnamefont{J.}~\bibnamefont{Horno}},
  \bibinfo{journal}{J. Colloid Interface Sci.} \textbf{\bibinfo{volume}{316}},
  \bibinfo{pages}{196} (\bibinfo{year}{2007}).

\bibitem[{\citenamefont{L\'opez-Garc\'ia
  et~al.}(2008)\citenamefont{L\'opez-Garc\'ia, Aranda-Rasc\'on, and
  Horno}}]{Lopez-Garcia2008}
\bibinfo{author}{\bibfnamefont{J.}~\bibnamefont{L\'opez-Garc\'ia}},
  \bibinfo{author}{\bibfnamefont{M.}~\bibnamefont{Aranda-Rasc\'on}},
  \bibnamefont{and} \bibinfo{author}{\bibfnamefont{J.}~\bibnamefont{Horno}},
  \bibinfo{journal}{J. Colloid Interface Sci.} \textbf{\bibinfo{volume}{323}},
  \bibinfo{pages}{146} (\bibinfo{year}{2008}).

\bibitem[{\citenamefont{Zhou and Schmid}(2012)}]{2012_ac}
\bibinfo{author}{\bibfnamefont{J.}~\bibnamefont{Zhou}} \bibnamefont{and}
  \bibinfo{author}{\bibfnamefont{F.}~\bibnamefont{Schmid}},
  \bibinfo{journal}{J. Phys.: Condens. Matter} \textbf{\bibinfo{volume}{24}},
  \bibinfo{pages}{464112} (\bibinfo{year}{2012}).

\bibitem[{\citenamefont{Zhou and Schmid}(2013)}]{2013_q0}
\bibinfo{author}{\bibfnamefont{J.}~\bibnamefont{Zhou}} \bibnamefont{and}
  \bibinfo{author}{\bibfnamefont{F.}~\bibnamefont{Schmid}},
  \bibinfo{journal}{Eur. Phys. J. E} \textbf{\bibinfo{volume}{36}},
  \bibinfo{pages}{33} (\bibinfo{year}{2013}).

\bibitem[{\citenamefont{Zhou et~al.}(2013)\citenamefont{Zhou, Schmitz,
  D\"unweg, and Schmid}}]{2013_response}
\bibinfo{author}{\bibfnamefont{J.}~\bibnamefont{Zhou}},
  \bibinfo{author}{\bibfnamefont{R.}~\bibnamefont{Schmitz}},
  \bibinfo{author}{\bibfnamefont{B.}~\bibnamefont{D\"unweg}}, \bibnamefont{and}
  \bibinfo{author}{\bibfnamefont{F.}~\bibnamefont{Schmid}},
  \bibinfo{journal}{J. Chem. Phys.} \textbf{\bibinfo{volume}{139}},
  \bibinfo{pages}{024901} (\bibinfo{year}{2013}).

\end{thebibliography}

\end{document}